\documentclass[prl,aps,twocolumn,floats,showpacspsfig]{revtex4}
\usepackage{amssymb}
\usepackage{latexsym}
\usepackage{amsmath}
\usepackage{epsfig}
\usepackage{color}
\usepackage{graphicx}
\usepackage{hyperref}

\newcommand{\be}{\begin{equation}}
\newcommand{\ee}{\end{equation}}
\newcommand{\bea}{\begin{eqnarray}}
\newcommand{\eea}{\end{eqnarray}}

\newcommand{\p}{\partial}
\newcommand{\s}{\sigma}

\newcommand{\la}{\langle}
\newcommand{\ra}{\rangle}
\newcommand{\rd}{\mbox{d}}
\newcommand{\ri}{i}
\newcommand{\re}{\mbox{e}}

\begin{document}
\title{Lattice spin models for  non-Abelian Chiral Spin Liquids}
\author{P. Lecheminant$^1$ and A.M. Tsvelik$^2$}
\affiliation{$^1$ Laboratoire de Physique Th\'eorique et
 Mod\'elisation, CNRS UMR 8089, Universit\'e de Cergy-Pontoise, Site de 
Saint-Martin,
F-95300 Cergy-Pontoise Cedex, France,\\
 $^2$ Division of Condensed Matter Physics and Materials Science, Brookhaven National Laboratory,
  Upton, NY 11973-5000, USA} \date{\today } 
\begin{abstract} 
We suggest a class of two-dimensional lattice spin Hamiltonians describing non-Abelian SU(2) chiral spin liquids - spin-analogues of fractional non-Abelian quantum Hall states-  with gapped bulk and gapless chiral edge excitations described by the SU(2)$_n$ Wess-Zumino-Novikov-Witten conformal field theory. The models are constructed
from an array of a generalized spin-$n/2$ ladders with multi-spin exchange interaction which are coupled by isolated spins.
Such models allow a controllable analytic treatment starting from the one-dimensional limit and are characterized by
a bulk gap and non-Abelian SU(2)$_n$ gapless edge excitations.

\end{abstract}


\maketitle

About thirty years ago Kalmeyer and Laughlin proposed a spin-analogue of a fractional quantum Hall state (FQHS) for electrically neutral quasiparticles - dubbed Kalmeyer-Laughlin chiral spin liquid (KL CSL) state  \cite{kl}. This state  shares basic properties of FQHS  such as gapped bulk and robust gapless edge excitations \cite{wwz,wen}.  The latter ones are  described by the SU(2)$_1$ Wess-Zumino-Novikov-Witten (WZNW)  conformal field theory \cite{balatsky,fradkin,wen}. The KL CSL state is a singlet and  breaks both time-reversal and parity symmetry. It is also topologically nontrivial with a vacuum degeneracy
on compactified spaces \cite{wen2}.

Historically, such a state was discovered by considering lattice versions of
the bosonic $\nu=1/2$ Laughlin wave function as variational candidate
ground-state wave functions for the triangular lattice spin-1/2 Heisenberg model \cite{kl,wwz,girvin}.
There is now a numerical evidence for its existence in  the spin-1/2 Heisenberg model on kagome lattice with the first, second and third nearest neighbor interaction \cite{chen1,chen2,gong}. Another piece of numerical evidence suggests that KL CLS exists for the spin-1/2 Heisenberg model on kagome and triangular lattices \cite{ludwig,lauchli1,lauchli2,chen3} with an additional three-spin exchange interaction explicitly breaking the time-reversal symmetry.  Both studies numerically demonstrate the required ground-state degeneracy and the full characterization of the underlying topological order for the KL CSL state.

Since numerics has its restrictions it is highly desirable to have microscopic models of CSLs which would allow a controlled analytical description. At the same time such controllable models should not be too unrealistic. While exact parent lattice Hamiltonians for KL CSL wave function have been constructed in Refs. \onlinecite{kivelson,greiterthomale,ciracsierra}, they are very challeging to achieve experimentally since they contain complex variables as well as long-range interactions.
A complementary approach is the so-called coupled-wires construction to FQHS  which starts from 
an array of one-dimensional (1D) wires, coupled in such a way that a two-dimensional (2D) gapped phase with
1D gapless edge excitations emerges \cite{teo}.
It opens a possibility to find microscopic models which display Abelian and non-Abelian topologically ordered physics
from arrays of quantum spin chains \cite{sela,thomale, fuji,mudry}.
For KL CSL such attempt was made by Gorohovsky {\it et.al.} \cite{sela} who suggested a model on anisotropic triangular lattice. Unfortunately the authors have found  that in any realistic situation the CSL will be destroyed by  competing magnetic orders. Since this puts them in a conflict with the numerical results of Refs. \onlinecite{lauchli2,chen3} who presented a good evidence for CSL even for the isotropic lattice, things are not under control. 

As was suggested in \cite{ronny,greiter2},  the KL CLS is just one example of spin liquid among many
and  CSLs with non-Abelian statistics might be stabilized.
We construct analytically tractable 2D microscopic models of non-Abelian SU(2) CSLs where the edge modes are described  
by the SU(2)$_n$ WZNW theory. For $n=2$ this is an analogue of the bosonic Pfaffian state.
For general $n$ these CSLs correspond to  the Read-Rezayi series at filling $\nu = n/2$ with  $\mathbb{Z}_n$ parafermionic neutral states  \cite{bosonic}. 
For $n \ge 2$, these quasiparticle excitations possess non-Abelian statistics like those of the Pfaffian.
The lattice models consist  of an array of coupled spin-$n/2$  generalized zigzag  ladders with three-spin exchange interactions which breaks the time-reversal and parity symmetries preserving their product. Although it looks unusual, such interaction may be generated  either as a result of spontaneous symmetry breaking \cite{wwz,baskaran} or, as was suggested in Ref. \onlinecite{chitra}, emerges in a Mott insulator as a result of a magnetic field.

{\bf Lattice spin models.}
The simplest lattice model for $n=1$  is depicted in Fig. 1 
and the Hamiltonian for individual ladders in this case includes only two- and three-spin interactions:
\bea
&& {\cal H}_{\rm ladder} = {\cal H}_J + {\cal H}_{\chi}, \label{model}\\
&& {\cal H}_J = \sum_{n}\Big[ J{\bf S}_n \cdot {\bf S}_{n+2} +J'{\bf S}_n \cdot{\bf S}_{n+1}\Big], \\
&& {\cal H}_{\chi} = \frac{\chi}{2}\sum_{i,j,k \in \Delta}\Big({\bf S}_i \cdot [{\bf S}_j\times{\bf S}_k]\Big),
\eea
where  the diagonal exchange is weaker then  the exchange along the legs: $ J' < J$. For the three-spin interaction the sum goes on all triangular plaquettes and indices $i,j,k$ appear in the clockwise order in each elementary triangle. 
\begin{figure}
\centerline{\includegraphics[angle = 0,
width=1.0\columnwidth]{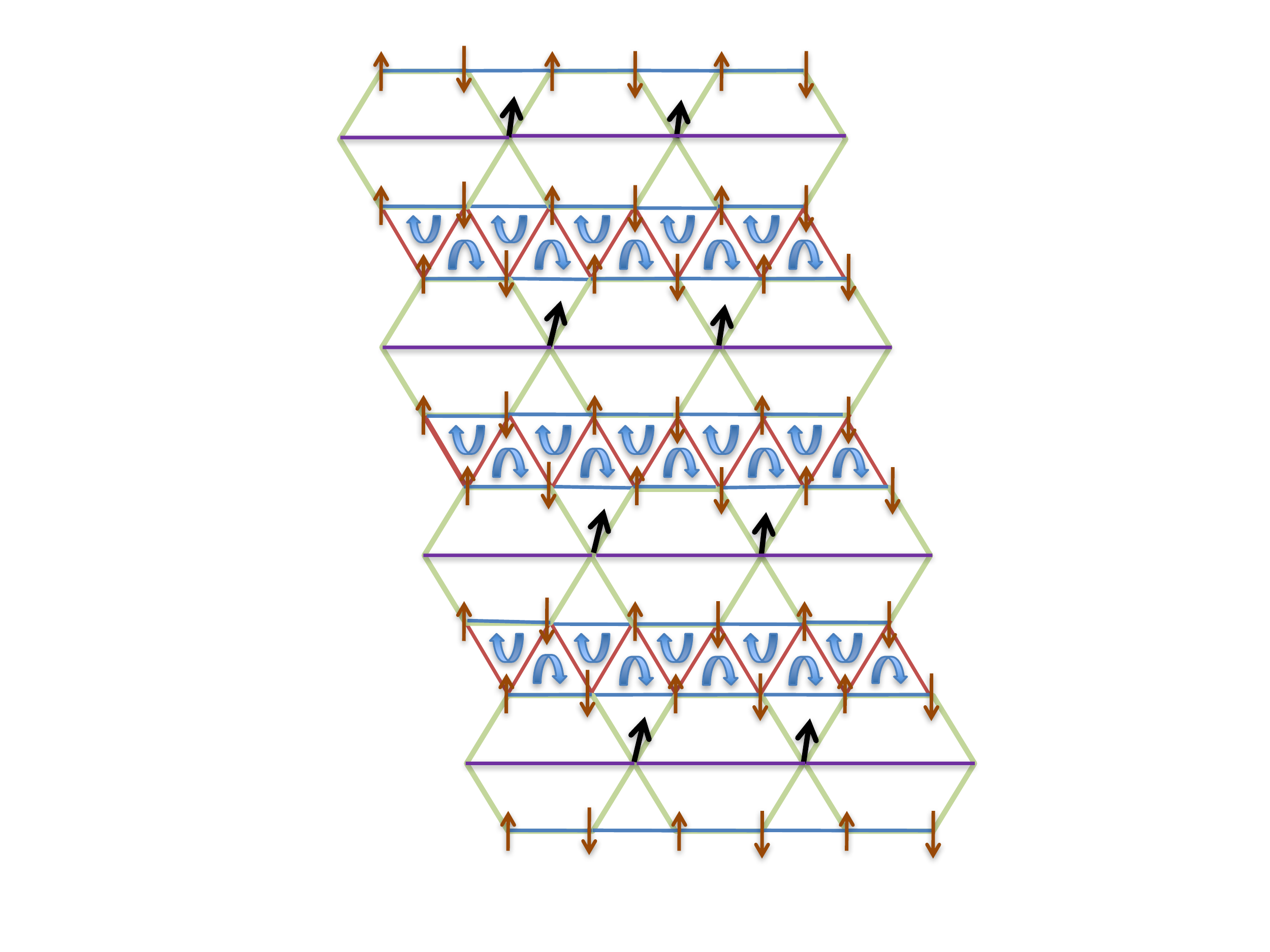}}
\caption{The spin arrangement for the suggested model. Different link colors correspond to different exchange interactions. The blue links mark the strongest isotropic exchange $J$, the red ones mark weaker isotropic exchange $J'$, the green links correspond to the anisotropic interladder exchange and the magenta ones to the Ising-like anisotropic exchange $I^{ab}$ between the intermediate spins. The blue arrows denote the three-spin exchange interactions. Since there is no static spin order all spin orientations depicted on the figure should be ignored.}
 \label{pict1}
\end{figure}
The ladder model (\ref{model})  allows analytic and controllable treatment in two cases. Either the interchain exchange is 
weak $J' << J$ \cite{sela} or one is close to the integrable point: $\chi^2 = 2JJ'$ \cite{frahm}.
The low-energy spectrum of the $n=1$ model (\ref{model}) consists of gapless bosonic modes of opposite chirality located on different legs of the ladder\cite{sela,frahm}. This  fulfills the necessary condition for the coupled-wires construction which requires that neighboring wires are coupled by chiral modes of opposite chirality \cite{teo}.
The authors of Ref. \onlinecite{sela} proceeded suggesting to couple the ladders in the fashion of  triangular lattice 
with the conclusion that such coupling will destroy  the CSL due to  the competing magnetic order. This competition comes from the coupling of staggered magnetizations of the next-to-nearest neighbor chains generated by higher order virtual processes. On the triangular lattice this interaction is $J_{nnn} \sim J''(J'/J)$, where $J''$ is the exchange between the ladders. On the other hand, the interaction between the gapless modes of the neighboring ladders is marginally relevant and the corresponding gap is exponentially small in $1/J''$ and hence  will be suppressed by the indirect exchange $J_{nnn}$.

In model of Fig. 1 the ladders with three-spin interactions which couple spins  around triangles are separated by localized spins. The interaction with the latter spins  is strongly anisotropic. Such  arrangement allows one to  overcome the problems related to the unfrustrated  interactions  between next-to-nearest spins discussed above and to make the CSL more robust, still providing  a possibility for analytical treatment of the model Hamiltonians. The fact that our model is formulated on a lattice allows numerical checks of our results. 

In the case of generic $n$, the integrable model (\ref{model}) for spins $S =n/2$ was found by Zvyagin \cite{zvyagin}. The explicit form of the Hamiltonian is given in Refs. \onlinecite{zvyagin,supp} and it represents coupled integrable critical spin-$S =n/2$ Heisenberg chains with extra multi-spin interactions around triangular plaquettes. The most important difference is that the gapless chiral modes now are described by the SU(2)$_n$ WZNW model. A 2D lattice model of non-Abelian CSL with  SU(2)$_n$ edge modes is constructed  then by coupling the generalized ladders as in Fig. 1.

{\bf The continuum limit of the individual ladder model.} 
To derive the effective field theory describing the continuum limit of the ladder model one just follows the standard approach 
of weakly-coupled two-leg spin ladders \cite{wznw,wznw1}. Alternatively at the integrable point one may use the Bethe ansatz results of Refs. \onlinecite{frahm,zvyagin}. The continuum limit Hamiltonian is the sum of two perturbed SU(2)$_n$ WZNW models with densities:
\bea
&& {\cal H}_{\rm eff} = {\cal H}_R +{\cal H}_L + {\cal V} \label{cont}\\
&& {\cal H}_R = \frac{2\pi v}{n+2}\Big[:{\bf J}_{R1}^2: + :{\bf J}_{L2}^2:\Big] + 2\pi v g_R {\bf J}_{R1}  \cdot {\bf J}_{L2}   \label{WZWR} \\
&& {\cal H}_L = \frac{2\pi v}{n+2}\Big[:{\bf J}_{L1}^2: + :{\bf J}_{R2}^2:\Big] + 2\pi v g_L {\bf J}_{L1}  \cdot {\bf J}_{R2}
\label{WZWL} \\
&&  {\cal V} = \gamma_{tw} {\bf n}_1  \cdot 
\overset{\leftrightarrow}{\partial_x} {\bf n}_2 + \gamma_{bs} ({\bf J}_{R1} \cdot {\bf J}_{L1}  + {\bf J}_{R2}  \cdot {\bf J}_{L2}), 
\eea
where ${\bf J}_{R,L}$ are SU(2)$_n$ Kac-Moody chiral currents and $v \sim J$ 
is the spin velocity. Operators ${\bf n}_{1,2}$ are staggered magnetizations of the chains 1,2. They have scaling dimension $d= 3/2(n+2)$ and are related to the $2\times 2$ matrix fields $G_{1,2}$ of the WZNW models describing individual chains\cite{wznw,wznw1}: ${\bf n}_{1,2} \sim \ri\mbox{Tr}[\vec\s G _{1,2}]$ ($\vec\s$ being a vector formed by the Pauli matrices).  For $n=1$ all perturbations are marginal while for $n >1$ the non-zero conformal spin perturbation with coupling constant $\gamma_{tw}$, dubbed twist term in Ref. \onlinecite{shura}, is strongly relevant with a scaling dimension $d_t = 1 + 3/(n+2)$. However, when the integrability condition is met, (i) $\gamma_{tw} =0, \gamma_{bs} < 0$ and the continuum limit Hamiltonian (\ref{cont}) splits into two independent parts and (ii) $g_R$ and $g_L$ have different signs. As a consequence of (ii) at the integrable point  one of the models ${\cal H}_{R,L}$ becomes massive and the other one is massless depending of the sign of $\chi$ with the marginally relevant dimensionless coupling $g = |g_R|  \sim (J'/2J)^{1/2}$. For $g << 1$ the spectral gap for massive mode at the integrable point is exponentially small:
$\Delta_{\rm ladder} = J g^{1/n}\exp(-1/g)$.
However, the gapless mode remains robust independently of the value of $g$ and hence always admits a field-theory description even when the gap is of the order of $J$. The field theory for the gapless mode  is the one of the SU(2)$_n$ WZNW models (\ref{WZWR}, \ref{WZWL}) which has a negative coupling constant. 

 According to Ref. \onlinecite{sela} who analyzed the one-loop renormalization group flows of model (\ref{cont}) with $n=1$ outside of the integrable point, the picture of the spectrum given above remains qualitatively valid in some region  outside of the integrable line. As a result  one can be confident that the phase diagram of the generalized zigzag spin ladder contains a region where the right movers of (say) chain 1 are strongly coupled to the left movers of chain 2 and the left movers of chain 1 are weakly coupled to the right movers of chain 2. The part of the spectrum containing the strongly coupled modes is gapped. Such violation of parity is, of course, a consequence of the three-spin interaction. For $n>1$ the situation is more delicate since the twist operator is relevant. Even if  the gap is sufficiently large so that  $\Delta_{\rm ladder} > \gamma_{tw}^{1/(2-d_t)}$, we have to make sure that the twist operator does not ruin the quantum criticality of the edge modes. In this respect, we need  to consider its projection  on the low-energy sector. Since the twist operator is a spin singlet and is a sum of two SU(2) singlet operators with Lorentz spin $\pm 1$, the projection must generate operators with the same symmetry or nothing.
  As described in Ref. \onlinecite{supp}, we find that the twist operator reduces to a total derivative of the trace of the SU(2)$_n$ WZNW  field (see Eq. (\ref{U}) below). It does not contribute to the effective low energy theory.

 The generalized zigzag ladder  is a critical spin nematic described  by one of the  SU(2)$_n$ WZNW models (\ref{WZWR}, \ref{WZWL}), the one which has negative coupling constant. The SU(2) WZNW matrix field $U$ of this model is made from  chiral components of the fields of the different chains \cite{supp}. 
In this theory the local operators with power law correlation functions are binary products of staggered magnetizations  from the different legs:
\bea
{\bf n}_1 \cdot {\bf n}_2 \sim \mbox{Tr} \; U, ~~ {\bf n}_1\times {\bf n}_2 \sim \ri\mbox{Tr}[\vec\s U ], \label{U}
\eea
while  two-point correlators of the staggered magnetizations decay exponentially \cite{supp}.
 In what follows we will assume that the parameters of the lattice Hamiltonian (\ref{model}) are such that the effective couplings for the current-current interactions on a given ladder  have different signs for different parities, as was described above.  

{\bf Coupling the ladders.} When zigzag ladders are arranged in a 2D array, the critical nematic modes couple. Depending on the nature of this coupling the system may either order as a nematic or remain a CSL. The order prevails if the matrix order parameters $U_y$ from different ladders couple directly. As discussed above, one needs to isolate the ladders from each other more thoroughly to stabilize a CSL state.  The arrangement shown  on  Fig. 1 
makes it more difficult for the nematic order parameters to couple. In that case the chiral gapless modes from neighboring ladders $y,y+1$ interact indirectly with each other through the intermediate spins interacting with a ferromagnetic Ising-like interaction. We will demonstrate that this arrangement produces  a spin gap in the bulk. As will be shown the gap magnitude increases  when the  exchange with the intermediate spins is anisotropic. Meanwhile  the exchange within the ladders must remain   SU(2) invariant to preserve the Kac-Moody algebra of the currents.  
Taking into account that each leg of an individual ladder contains just one chiral gapless mode
and that the in-chain staggered magnetizations are short-ranged fields at energies $ \ll \Delta_{\rm ladder}$, 
we arrive to the following low-energy model which describes the neighboring ladders $y,y+1$ as a 
 Kondo lattice model with a forward scattering:
\bea
&& {\cal H} = \frac{2\pi v}{n+2} \int \rd x \Big[:{\bf J}_{R,y+1}^2: + :{\bf J}_{L,y}^2:\Big] 
\nonumber \\
&+& \sum_{l}  \eta_a  \Big[ J^a_{L,y}(x_l) +  J^a_{R,y+1}(x_l)\Big]S^a_{y+1/2}(x_l) \label{Kondo}\\
&-&   \sum_{l} I^{ab}S^a_{y+1/2}(x_l)S^b_{y+1/2}(x_{l+1}). \nonumber
\eea
We will consider the case $\eta^z >> \eta^x =\eta^y$. According to Ref. \cite{oleg} where somewhat similar system  was studied, Hamiltonian (\ref{Kondo}) must be augmented by the RKKY interaction between the localized spins generated by the coupling to irrelevant operators. The strongest of those is $\eta_z\p_x (n^z_y+ n^z_{y+1})S^z_{y+1/2}$ in 
our anisotropic model. Since correlation functions of the staggered magnetization decay exponentially, the second-order perturbation theory in $\eta_a$ yields the short range interaction ${\cal J}_r^{ab}S^a_nS^b_{n+r}$, where the exchange ${\cal I}_r$ decays exponentially with distance \cite{supp}. This interaction is antiferromagnetic and competes with the bare one. We will consider the case when the net interaction $I^{ab}_r = -I^{ab} \delta_{r,1} + {\cal J}^{ab}_r$ is small. 
For $n=1$ model (\ref{Kondo}) with $I_r=0$ was considered in Refs. \onlinecite{Emery,coleman} and it was concluded that the spectrum is gapped and has a nonlocal order parameter.  
Using the Abelian bosonization approach, one can  bosonize the chiral SU(2)$_1$ currents in terms of chiral bosons $\Phi_{R,L}$ and  recast the Lagrangian density for model (\ref{Kondo}) as (from now on we set $v=1$)
\bea
 {\cal L} &=&  \frac{1}{2}[(\p_{\tau}\Phi)^2 +(\p_x\Phi)^2] + \rho_s\Big\{\frac{\ri}{2}S^z\p_{\tau}\psi + \frac{\eta_z}{\sqrt{2\pi}}\p_x\Phi S^z  \nonumber \\
 &+& \frac{\eta_{\perp}}{2\pi a_0}\cos(\sqrt{2\pi}\Phi)(S^+\re^{-\ri\sqrt{2\pi}\Theta}  +
 S^-\re^{\ri\sqrt{2\pi}\Theta})\Big\} \nonumber\\
 & +&  \sum_{n,r}  I^{ab}_{r} S^a_n S^b_{n+r}\label{ZEK},
\eea
where we assume that the local spins can be located at arbitrary points along the chains  with spin density $\rho_s$, and not necessarily in a regular fashion as depicted on Fig. \ref{pict1}. Here $\Theta = \Phi_{L} - \Phi_{R}$ 
is the field dual to $\Phi = \Phi_{L} + \Phi_{R}$ and $S^{\pm} = \re^{\pm\ri\psi}\sin\theta, ~~S^z=\cos\theta$. After the transformation $\psi \rightarrow \psi +\sqrt{2\pi}\Theta$ the coupling in the second term changes to $\eta_z \rightarrow \eta_z - 2\pi$ and the Hamiltonian becomes
\bea
&& {\cal H} =  \sum_{n,r}  I^{ab}_{r} \tilde S^a_n\tilde S^b_{n+r}  
+ \eta_{\perp} \sum_n  \tilde S^x_n\cos(\sqrt{2\pi}\Phi_n)  \label{ZEK1} \\
&+&\frac{\eta_z-2\pi}{\sqrt{2\pi}} \sum_n \p_x  \Phi_n \tilde S_n^z + \frac{1}{2} \int \rd x [(\p_x\Theta)^2 + (\p_x\Phi)^2], \nonumber
\eea
The situation simplifies when  $\eta^z=2\pi$. As we have said, we we will always consider the case when $I_r$ is sufficiently small which can be always achieved by tuning the interactions.  In the simplest case $I_r =0$ we are left  with the interaction term
\be
V =  \frac{\eta_{\perp}\rho_s}{2\pi a_0}\cos(\sqrt{2\pi}\Phi)\tilde S^x,
\label{SG}
\ee
where $\tilde S^a$ are transformed spin components which includes the staggered factor \cite{Emery,coleman}.
 Then since the operators $\tilde S^x(x)$ commute with the Hamiltonian, they can be replaced by constants. The ground state configuration corresponds to all $\tilde S^x(x)$ being equal. In this sector model (\ref{SG}) is reduced to the integrable sine-Gordon model with a spectral gap $\Delta \sim (\rho_s\eta_{\perp})^{2/3}$. The dimensional analysis indicates that for general $\eta_z >> \eta_{\perp}$ the gap is $\Delta \sim \eta_{\perp}^{4\pi/3\eta_z}.$
So the bulk excitations are gapped and the boundary chiral modes are gapless as is expected for a FQHS. The latter ones are described by the SU(2)$_1$ WZNW theory.  A similar Toulouse approach can be performed in the general $n>1$ case,
as described in Ref. \onlinecite{supp}, with the emergence of a bulk gapped phase with gapless edge excitations
described by the SU(2)$_n$ WZNW conformal field theory. 

 The  gap $\Delta$ provides us with the window of tolerance for the magnitude of $I_r$: this exchange interaction should not exceed $\Delta$. We find it interesting to take a closer look at the bulk excitations in the $n=1$ case. At the intergrable point $\eta_z = 2\pi, I_r =0$  model (\ref{ZEK1}) has two kinds of excitations: the massive mobile sine-Gordon triplet modes and immobile defects corresponding to flips of $\tilde S^x$. Away from this  point the latter defects   become mobile solitons; each soliton carries spin 1/2 from the ladders and a zero Majorana mode from the interladder spins. To see this we need to consider a perturbation around $\eta^z=2\pi$. Integrating out the massive $\Phi$ modes  in the leading order in  $\delta\eta = \eta_z-2\pi$ changes $I_r$ by the amount  
$\delta I_{r} \sim (\delta\eta/v)^2(\Delta/v)\exp(-\Delta |r|/v)$. Treating the exponentially decaying Ising exchange as the nearest neighbor interaction we arrive to the Ising model in a transverse dynamical field. The Jordan-Wigner transformation brings us to the model of Majorana fermions which mass term changes sign when $\cos(\sqrt{2\pi}\Phi)$ does. Such model has been considered in Ref. \onlinecite{wild} and in the limit when the Majoranas are slower than kinks possesses rich physics. It certainly has Majorana zero modes riding on kinks. There are also traces of the two-channel Kondo physics suggested for this case in Ref. \onlinecite{ludwig} in the sense that the Majorana zero modes on kinks are similar to static Majorana zero modes on isolated spins in the two-channel Kondo effect \cite{ludwigPr}. So we see that the bulk of our model  is gapped and the excitations are nontrivial. A half of the spinon modes of the ladders are gapped inside of each ladder. These are spin-1/2 solitons. The remaining half are gapped due to the interladder coupling through the intermediate spins. These are non-Abelian kinks (visons?). Such fractionalized excitations cannot propagate in the transverse direction. However, there are probably  their bound states with integer quantum numbers and these ones  can. If the applied magnetic field exceeds the bulk gap we  expect the bulk to become gapless with central charge $c=3/2$ and the velocity $v \sim M$ (the magnetization). This may lead to interesting $T/M$ scaling in the thermodynamics. 

 In order to demonstrate the robustness of the described state we have to consider a possibility of interladder couplings other than the ones already described.  In the low-energy theory such couplings are generated by high energy virtual processes and their magnitudes are proportional to powers of the couplings of the bare lattice Hamiltonian \cite{starykh}. In the given case one has to expect, for instance, the  appearance of the effective interactions between the local (or Kondo) spins from different rows which may lead to their ordering. This process  competes  with the Kondo screening preventing the gap formation. Such unwanted interactions constitute a general problem for all quantum Hall constructions based on coupling of wires. Starting from the original idea by Teo and Kane \cite{teo}  all these scenaria envisage that  the original  lattice model can at low energy be reduced to 1D critical models coupled by marginally relevant current-current interactions. The difficulty is that the gaps generated by the marginal interactions are exponentially small  in the {\it direct} interwire exchange and the ones generated by high energy virtual processes have power law dependence on the {\it same}  couplings of the lattice model. 
In the model with the intermediate spin we suggest this difficulty is avoided since the spin currents from neighboring ladders are not coupled directly which would generate exponentially small gap, but via the anisotropic exchange with intermediate spins.

\begin{acknowledgements} 
We are grateful to C. Chamon, F. H. L. Essler, Y. Fuji, V. Gritsev, R. M. Konik,  N. Robinson, O. A. Starykh and A. Weichselbaum for inspirational discussions. A. M. T. thanks the Isaac Newton Institute in Cambridge for hospitality. The authors acknowledge the Yukawa Institute for Theoretical Physics for hospitality during the completion
of this work.  A.M.T. was supported by the US DOE under contract number DE-AC02-98 CH 10886.
P. L. would like to thank CNRS (France) for financial support (PICS grant).

\end{acknowledgements}

\newpage
\widetext
\begin{center}
\textbf{\large Supplemental Materials: Lattice spin models for  non-Abelian Chiral Spin Liquids}
\end{center}
\setcounter{equation}{0}
\setcounter{figure}{0}
\setcounter{table}{0}
\setcounter{page}{1}
\makeatletter
\renewcommand{\theequation}{S\arabic{equation}}
\renewcommand{\thefigure}{S\arabic{figure}}
\renewcommand{\bibnumfmt}[1]{[S#1]}
\renewcommand{\citenumfont}[1]{S#1}
\makeatother

\section{The explicit form of the integrable spin-$S$  zigzag two-leg ladder}

The integrable spin-$S$ two-leg zigzag ladder with a time-reversal breaking interaction is defined 
by the lattice Hamiltonian  \cite{zvyaginapp}: 
\bea
&& {\cal H} = \sum_n\Big[\theta^2({\cal H}_{n_1,n_1+1} + {\cal H}_{n_2,n_2+1} ) +2({\cal H}_{n_1,n_2} + {\cal H}_{n_1,n_2+1} )+ \nonumber\\
&& \{({\cal H}_{n_1,n_1+1} + {\cal H}_{n_2,n_2+1} ),({\cal H}_{n_1,n_2} + {\cal H}_{n_1,n_2+1} )\} + \nonumber\\
&& 2\ri\theta[({\cal H}_{n_1,n_1+1} + {\cal H}_{n_2,n_2+1} ),({\cal H}_{n_1,n_2} + {\cal H}_{n_1,n_2+1} )]\Big],
\label{intmodapp}
\eea
where 
\bea
&& {\cal H}_{n,n+1}(x) = \sum_{j=1}^{2S}\sum_{k=1}^j\frac{k}{k^2+\theta^2}\times \prod_{l=0, l \ne j}^{2S}\frac{2x - l(l+1) + 2S(S+1)}{j(j+1)-l(l+1)},\nonumber\\
&& x = {\bf S}_{n} \cdot {\bf S}_{n+1}.
\eea
When $\theta=0$, the latter is the integrable spin-$S$ Heisenberg chain \cite{BTapp} 
which belongs to SU(2)$_{2S}$ universality class \cite{affleckapp}.
An explicit form of model (\ref{intmodapp}) for $S=1$ can be found in Ref. \cite{tavaresapp}.
 The Bethe ansatz equations for model (\ref{intmodapp}) are \cite{zvyaginapp,frahmapp}: 
 \bea
 && \Big(\frac{\lambda_a - \theta + \ri n/2}{\lambda_a - \theta - \ri n/2}\Big)^N\Big(\frac{\lambda_a + \theta + \ri n/2}{\lambda_a + \theta - \ri n/2}\Big)^N = \prod_{b=1}^M\frac{\lambda_a - \lambda_b + \ri}{\lambda_a - \lambda_b - \ri}, \\
 && E = \sum_{a=1}\sum_{\s = \pm 1} \frac{n}{(\lambda_a + \s\theta)^2 + n^2/4} .
\eea

\section{Nematic order parameters of the individual ladder model}

We consider a single generalized zigzag two-leg spin ladder where the sign of the T-breaking term is such that  for instance $g_L>0$ and $g_R <0$. This means that the $2R$ and $1L$ modes hybridize and are gapful. In stark contrast,
the other chiral modes $2L$ and $1R$ remain gapless and correspond to the edge modes with quantum
critical behavior. 
Let us try to write the low-energy limit of various order parameters after averaging out the massive
degrees of freedom as it is done in \cite{CO}.  We first consider the scalar product of staggered magnetizations:
\bea
{\bf n}_1 \cdot {\bf n}_2 &\sim& - \langle \mbox{Tr}[\vec\s G_1]  \mbox{Tr}[\vec\s G_2]\rangle
 \nonumber \\
&\sim& \langle  -2  \mbox{Tr} (G_1 G_2) +  \mbox{Tr} (G_1)   \mbox{Tr} (G_2) \rangle
 \nonumber \\
 &\sim& -2  \langle G_{1L \alpha} G_{2 R \alpha}   \rangle G_{2L \beta} G_{1 R \beta}
 +    \langle G_{1L \alpha} G_{2 R \beta}   \rangle G_{2L \beta} G_{1 R \alpha},
 \label{nematicOP}
\eea
where $\langle A\rangle$ means the average of operator A in the ground state of the massive theory of ${\cal H}_L$ 
and we have separated the chiral components of the SU(2)$_n$ WZNW matrices  $G_{1,2}$ ($\alpha,\beta=
\uparrow, \downarrow$).
In the ground state of the current-current model ${\cal H}_L$, we have: 
 $\langle G_{1L \alpha} G_{2 R \beta}   \rangle = C \delta_{\alpha \beta}$, $C$ being a non-universal constant.
 By introducing the emerging  SU(2)$_n$ WZNW matrix $U_{\alpha \beta} = G_{2L \alpha} G_{1 R \beta}$,
 we find the low-energy description from Eq. (\ref{nematicOP}): ${\bf n}_1 \cdot {\bf n}_2 \sim \mbox{Tr}  \; U$.
 A similar approach can be perfomed for the vector chiral order parameter:
 ${\bf n}_1 \wedge {\bf n}_2 \sim i \mbox{Tr} ( \; \vec\s   U)$. Finally, the staggered magnetization of each individual chain
 ${\bf n}_{1,2}$ has a zero vacuum expectation value in the ground state of model ${\cal H}_L$
 which means that their two-point correlators decay exponentially at large distance and at low-energy.

 We now discuss the fate of the twist term ${\bf n}_1  \cdot 
\overset{\leftrightarrow}{\partial_x} {\bf n}_2 $ after averaging out the massive
degrees of freedom of the integrable model ${\cal H}_L$. By performing a similar approach as in Eq. (\ref{nematicOP}),
we get: ${\bf n}_1  \cdot 
\overset{\leftrightarrow}{\partial_x} {\bf n}_2 \sim   G_{2L\alpha }
\overset{\leftrightarrow}{\partial_x}  G_{1 R \alpha}$. Using $\partial_x = i (\partial - {\bar \partial})$ and
$\partial_{\tau} = i (\partial + {\bar \partial})$, we find that the projection of the twist term in the low-energy limit
reduces to a total derivative of the SU(2)$_n$ WZNW field: ${\bf n}_1  \cdot 
\overset{\leftrightarrow}{\partial_x} {\bf n}_2 \sim \partial_{\tau} \mbox{Tr} U$ and can be thus neglected.

\section{The effective interactions}

 As is written in the main text, the exchange interactions between local spins and staggered magnetizations of the chains give rise to short range interactions between the latter spins. Below we will provide explicit calculations to support this statement. The interaction is generated by integration over massive modes encoded in staggered magnetization of the ladders. In the second order of perturbation theory in $\eta^a$ we get the following contribution to the action of the spins:
 \bea
\delta S =  -\frac{1}{2}\eta_a^2 \sum_{j,k}\int \rd\tau_1\rd\tau_2 S^a_{j+1/2}(\tau_1)\la\la (n_j -n_{j+1})^a(\tau_1)(n_k - n_{k+1})^a(\tau_2)\ra\ra S^a_{k+1/2}(\tau_2)
\eea
We can calculate the correlation functions explicitly. Due to the SU(2) symmetry of the ladder Hamiltonian it is sufficient to do it for $n^z \sim \sin(\sqrt{2\pi}\Phi) = \sin[\sqrt{2\pi}(\varphi + \bar\varphi)]$. We will take advantage of the fact that for a given chain $\varphi$ and $\bar\varphi$ do not couple to each other and one of these field is gapless and the other is gapped. Then we have 
\bea
&& \la\la \sin[\sqrt{2\pi}\Phi(1)] \sin[\sqrt{2\pi}\Phi(2)]\ra\ra = \frac{1}{2}\la\la \re^{\ri\sqrt{2\pi}\varphi(1)}\re^{-\ri\sqrt{2\pi}\varphi(2)}\ra\ra \la\la \re^{\ri\sqrt{2\pi}\bar\varphi(1)}\re^{-\ri\sqrt{2\pi}\bar\varphi(2)}\ra\ra \sim \nonumber\\
&& \frac{1}{(v\tau_{12}+ \ri x_{12})^{1/2}}D(1,2).
\eea
The correlation function $D(1,2)$ was calculated in \cite{Essler}:
\bea
D(\tau,x) \equiv \la\la \re^{\ri\sqrt{2\pi}\bar\varphi(\tau,x)}\re^{-\ri\sqrt{2\pi}\bar\varphi(0,0)}\ra\ra \approx Z \Big(\frac{v\tau +\ri x}{v\tau - \ri x}\Big)^{1/4} r^{-1/2}\re^{-\Delta_{ladder} r}, ~~ r^2 = \tau^2 + (x/v)^2,
\eea
where $Z\sim 1$, so that 
\bea
\la\la n^a(1)n^a(2)\ra\ra \sim r_{12}^{-1}\exp(-\Delta_{ladder} r_{12}).
\eea
Since $\eta^z >> \eta^{x,y}$ the most important contribution is to the $S^zS^z$ interaction as is stated in the main text. The Fourier transform of the exchange integral is given by $\omega =0$ transform of the correlation function: 
\bea
{\cal J}^{aa}(q) \sim -\frac{[\eta^a]^2q^2 }{\sqrt{(qv)^2 + \Delta^2_{ladder}}}\rightarrow \frac{[\eta^a]^2}{v}\p_x^2K_0(\Delta_{ladder}x/v)
\eea
This interaction is antiferromagnetic and decays fast in real space (as $1/x^2$) even at small distances. 

 The integration over the massive modes will also introduce interaction between localized spins of different chains. However, since the correlation function between $n_1$ and $n_2$ is zero, this interaction is generated  only in the fourth order in $\eta$:
 \bea
 S_{j+1/2}(y)S_{k+1/2}(y+1) S_{l+1/2}(y)S_{m+1/2}(y+1) \la\la \p_x n(y,x_1)\p_x n(y+1,x_2) \p_x n(y,x_3)\p_x n(y+1,x_4)\ra\ra 
\eea
$x_1\approx x_3, x_2 \approx x_4$ so this will generate the interaction of energy densities of the Ising models.

\section{Toulouse limit solution of the Kondo-lattice model with forward scattering}

We extend the Toulouse limit solution of the Kondo-lattice model with forward scattering presented in the letter to the $n \ge 2$ case. The Hamiltonian is defined by:
\bea
 {\cal H} = \frac{2\pi v}{n+2} \int \rd x \Big[:{\bf J}_{R,y+1}^2: + :{\bf J}_{L,y}^2:\Big] 
+ \sum_{l}  \eta_a  \Big[ J^a_{L,y}(x_l) +  J^a_{R,y+1}(x_l)\Big]S^a_{y+1/2}(x_l),  \label{Kondoapp}
\eea
where ${\bf J}_{R,y+1}$ and ${\bf J}_{L,y}$ are chiral SU(2)$_n$ Kac-Moody currents.
The starting point of the solution is to express these currents
in terms of Z$_N$ parafermions  currents  $\psi_L,\psi_R$ with conformal weights (1-1/n,0),  (0,1-1/n) and chiral bosonic fields $\Phi_{L,R}$ \cite{para}:
\bea
J_{L,y}^z &=& \sqrt{\frac{n}{2\pi}}\p_x \Phi_L, ~~J_{L,y}^{+}=\frac{\sqrt n}{2\pi a_0}\re^{\ri\sqrt{8\pi/n}\Phi_L}\psi^{+}_L  \nonumber \\
J_{R,y+1}^z &=& \sqrt{\frac{n}{2\pi}}\p_x \Phi_R, ~~J_{R,y+1}^{+}=\frac{\sqrt n}{2\pi a_0}\re^{-\ri\sqrt{8\pi/n}\Phi_R}\psi^{+}_R  .
\eea
For the sake of simplicity, we assume that the chiral fields with different chiralities commute between themselves.
The interacting part of model (9) reads then as follows:
\bea
 {\cal H}_{\rm int} =  \sum_{l}   \eta_z  \sqrt{\frac{n}{2\pi}}\p_x \Phi  S^z 
 + \frac{ \eta_{\perp} \sqrt n}{2\pi a_0} S^{+}  \re^{-\ri\sqrt{2\pi/n}\Theta}
 \left[  \re^{-\ri\sqrt{2\pi/n}\Phi} \psi_L +  \re^{\ri\sqrt{2\pi/n}\Phi} \psi_R \right] + {\rm H.c.},
 \label{Kondoint}
\eea
where $\Phi = \Phi_L + \Phi_R$ and $\Theta = \Phi_L - \Phi_R$  are respectively the total bosonic field
and its dual field. Repeating the same procedure as for $n=1$, we now absorb the phase factor
$\re^{-\ri\sqrt{2\pi/n}\Theta}$ into the spin operator by a canonical transformation:
$U^{+} S^{+}(x_l) U   =S^{+} (x_l) \re^{\ri\sqrt{2\pi/n}\Theta(x_l)}$.
The canonical transformation is then defined by:
\bea
U =  \re^{-\ri \sqrt{\frac{2\pi}{n}} \sum_l \Theta(x_l) S^{z}(x_l)}.
\eea
Since $\left[  \Theta(x), \Phi(y) \right] = -i \theta( x-y)$, we get the non-trivial transformation for the total bosonic field:
\bea
U^{+} \Phi (x) U &=&   \Phi (x)  + \sqrt{\frac{2\pi}{n}} \sum_l  S^{z}(x_l)\theta( x_l -x) 
\label{toulousetrans} \\
\frac{1}{2}  \int \rd x U^{+} (\p_x \Phi)^2 U &=&  \frac{1}{2}  \int \rd x  (\p_x \Phi)^2 
- \sqrt{\frac{2\pi}{n}} \sum_l  S^{z}(x_l)\p_x \Phi (x_l) .
\eea
The Toulouse limit solution is then defined when $\eta_{z}= 2\pi/n$: 
\bea
{\cal H}^{'}_{\rm int} = U^{+} {\cal H}_{\rm int} U =   \frac{\eta_{\perp} \sqrt{n}}{2\pi a_0}  \sum_l S^{+} (x_l)
 \left[  \re^{-\ri\sqrt{2\pi/n}\Phi} \psi_L +  \re^{\ri\sqrt{2\pi/n}\Phi} \psi_R \right](x_l) + {\rm H.c.},
\eea
where, in the parafermion fields,  we have absorbed phase factors which stem from the non-trivial
transformation of the $\Phi$ field (\ref{toulousetrans}) under the canonical transformation. The resulting interaction is strongly relevant and opens spectral gaps for the bulk modes.

\end{document}